\documentclass[12pt]{iopart}
\usepackage{amssymb,setstack,mathdots}

\def\ba{\begin{array}}
\def\ea{\end{array}}
\def\be{\begin{equation}}
\def\ee{\end{equation}}
\def\nn{\nonumber}

\def\sp{{\rm span}}

\def\C{{\mathbb C}}
\def\R{{\mathbb R}}
\def\N{{\mathbb N}}
\def\F{{\mathbb F}}

\def\b{{\mathfrak b}}

\def\g{{\mathfrak g}}
\def\h{{\mathfrak h}}

\def\n{{\mathfrak n}}
\def\r{{\mathfrak r}}
\def\p{{\mathfrak p}}

\def\s{{\mathfrak s}}
\def\glmot{{\mathfrak g \mathfrak l}}
\def\slmot{{\mathfrak s \mathfrak l}}
\def\so{{\mathfrak s \mathfrak o}}
\def\t{{\mathfrak t}}
\def\z{{\mathfrak z}}
\def\n{{\mathfrak n}}
\def\der{{\mathfrak{Der}}}
\def\inn{{\mathfrak{Inn}}}

\def\span{{\rm span}}

\def\NR{{\rm NR}}
\def\ad{{\rm ad}}

\def\rank{{\rm rank}}
\def\out{{\mathfrak{Out}}}

\newcommand{\sdir}{\ensuremath{\rlap{\raisebox{0.15ex}{$\mskip 6.5mu\scriptstyle+$}}\supset}}
\begin{document}

\newtheorem{theorem}{Theorem}
\newtheorem{prop}{Proposition}
\newtheorem{conjecture}{Conjecture}

\title[Solvable Lie algebras with Borel nilradicals]{Solvable Lie algebras with Borel nilradicals}

\author{L \v Snobl\dag\   and P Winternitz\ddag}

\address{\dag\ Faculty of Nuclear Sciences and Physical Engineering, 
Czech Technical University in Prague, B\v rehov\'a 7, 115 19 Prague 1, Czech Republic }

\address{\ddag\ Centre de recherches math\'ematiques and D\'epartement de math\'ematiques et de statistique, 
Universit\'e de Montr\'eal, CP 6128, Succ. Centre-Ville, Montr\'eal (Qu\'ebec) H3C 3J7, Canada}

\eads{\mailto{Libor.Snobl@fjfi.cvut.cz}, \mailto{wintern@crm.umontreal.ca}}

\begin{abstract}
The present article is part of a research program the aim of which is to find all indecomposable solvable extensions of a given class of nilpotent Lie algebras. Specifically in this article we consider a nilpotent Lie
algebra $\n$ that is isomorphic to the nilradical of the Borel subalgebra of a complex simple Lie algebra, or of its split real form. We treat all classical and exceptional simple Lie algebras in a uniform manner. We
identify the nilpotent Lie algebra $\n$ as the one consisting of all positive root spaces. We present general structural properties of all solvable extensions of $\n$. In particular, we study the extension by one nonnilpotent element and by the maximal number of such elements. We show that the extension of maximal dimension is always unique and isomorphic to the Borel subalgebra of the corresponding simple Lie algebra.
\end{abstract}



\section{Introduction}\label{intro}

In applications in physics, other natural sciences and engineering, Lie groups usually occur as groups of transformations, acting on certain manifolds and leaving certain objects invariant. Their Lie algebras are obtained as sets of matrices or operators, differential, difference or other. The first task that occurs is to identify the obtained Lie algebra as an abstract Lie algebra. This may be difficult because no complete classification of Lie algebras exists. The task is facilitated by Levi theorem \cite{Levi,Jac} which states that every finite dimensional Lie algebra $\g$ over a field of characteristic zero can be written as the semidirect sum of a semisimple Lie algebra $\p$ and a solvable one $\r$
\begin{equation}\label{levi}
\g=\p \sdir \r
\end{equation}
In equation (\ref{levi}) $\r$ is the radical of $\g$, i.e. its maximal solvable ideal and the semisimple Lie algebra $\p$ is called Levi factor of $\g$.

In this article we restrict the considered field to be either $\F=\C$ or $\F=\R$. Any semisimple Lie algebra is a direct sum of simple factors and all simple Lie algebras over $\C$ have been classified by Cartan \cite{Cartan} and over $\R$ by Cartan and by Gantmacher \cite{Gan}.

A lot of information on solvable and nilpotent Lie algebras is available in the literature \cite{Supr,Vergne,GKh2} but no complete classification of such algebras exists, nor can it exist.  Only very special classes of nilpotent Lie algebras were classified in an arbitrary finite dimension, e.g. naturally graded filiform \cite{Vergne} or naturally graded quasi--filiform \cite{GJM} ones. The classification of finite dimensional nilpotent Lie algebras exists for dimensions $n\leq 6$ \cite{Morozov} and $n=7$ \cite{Saf,Seeley,Gong}. In higher dimensions, i.e. $n=8,9$, some classifications of nilpotent algebras obtained using computer programs exist \cite{Tsa1,Tsa2}. Solvable Lie algebras were classified up to dimension $n=5$ in \cite{Mub1,Mub2} and for $n=6$ in \cite{Mub3,Tur2}. (The results of the papers \cite{Morozov} and \cite{Mub2}, which were written in Russian, are reviewed in English in \cite{PSW} where the corresponding Casimir invariants were computed.) Also  a partial classification, i.e. solvable algebras with 6--dimensional nilradicals, exists in dimension 7 \cite{Parry}. Some particular classes of nilpotent and solvable Lie algebras were classified up to slightly higher dimensions (e.g. up to $n=11$ for filiform algebras in \cite{GKh1,GJMKh}, up to $n=8$ for rigid solvable Lie algebras in \cite{GKh2}). Some results on the general structure of filiform Lie algebras in an arbitrary dimension are also available in \cite{GKh}.

Instead of proceeding by dimension in the classification of solvable Lie algebras, it is possible to proceed structurally. We notice that many nilpotent Lie algebras, called \emph{characteristically nilpotent}, cannot be nilradicals of any solvable Lie algebra and should be excluded from the consideration \cite{Kh,GKh}. Thus, we start from an infinite class of nilpotent, but not characteristically nilpotent, Lie algebras that is already known for all dimensions $n$. Some such series are immediately obvious, like Abelian Lie algebras, Heisenberg algebras, or Lie algebras of strictly upper triangular matrices. Other infinite series of nilpotent algebras can be obtained by taking algebras from existing lists of low dimensional nilpotent Lie algebras and generalizing them to arbitrarily higher dimensions. For a given but unspecified dimension $n$ an algebra in a chosen class of
nilpotent Lie algebras can serve as the nilradical of a set of solvable Lie algebras. These are then obtained as solvable extensions of the chosen nilpotent ones.

The nilpotent Lie algebras that have so far been treated as nilradicals in this manner are Heisenberg algebras \cite{RW}, Abelian algebras \cite{NW}, triangular Lie algebras \cite{TW}, certain classes of filiform Lie algebras \cite{SW,SW1,ACSV,CS} and some others \cite{WLD,SK,ACSV1}.

The purpose of this article is to consider a different infinite class of nilpotent Lie algebras that may be of special interest in physics because of their close relationship with simple Lie algebras. Indeed, every complex simple Lie algebra $\g$ has a uniquely defined (up to equivalence) maximal solvable subalgebra, its Borel subalgebra $\b(\g)$. In turn, the Borel subalgebra (like every solvable Lie algebra) has a unique maximal nilpotent ideal, its nilradical $NR(\b(\g))$. It is this class of nilpotent Lie algebras that we shall extend to solvable Lie algebras. Thus $NR(\b(\g))$ will be the nilradical of the obtained solvable Lie algebras. We shall call these nilpotent algebras ``Borel nilradicals''. For each simple complex Lie algebra $\g$ the Borel nilradical is the Lie algebra consisting of all positive root spaces. It is of course the nilradical only of the Borel subalgebra $\b(\g)$, not of $\g$ itself. 

For real Lie algebras the maximal solvable subalgebra is in general not unique. The maximal solvable subalgebras of all real classical algebras were classified in \cite{PWZsupq,PWZsopq,Perroud2}. For the split real form of the complex simple Lie algebra there exists a distinguished maximal solvable subalgebra that has the same basis as the Borel subalgebra but considered over $\R$ instead of $\C$. We shall also call their nilradicals  ``Borel nilradicals'' (over $\R$), sometimes adding the adjective ``split'' to highlight its relation to the split real form.

For the simple Lie algebras $A_l$ the Borel nilradical is isomorphic to the Lie algebra of all strictly upper triangular matrices in $\slmot(l+1,\F)$ for $\F=\C$ or $\F=\R$, respectively. The solvable extensions of these nilradicals were obtained in an earlier article \cite{TW}.

For all other series of classical complex Lie algebras and their split real forms the nilradicals are also isomorphic to Lie algebras of strictly upper triangular matrices in $\C^{n\times n}$ or $\R^{n\times n}$  satisfying an additional condition 
\begin{equation}\label{xkkxt}
X K + K X^t=0.
\end{equation}
For the algebras $B_l,C_l,D_l$ the matrix $K$ is chosen to be 
\begin{eqnarray}
\nonumber B_l  =  \so(2l+1,\C), \,\so(l+1,l) & \qquad &   K  =J_{2l+1},\\
 C_l =  \mathfrak{s}\mathfrak{p}(2l,\C), \,\mathfrak{s}\mathfrak{p}(2l,\R) & \qquad & K  =\left( \begin{array}{cc} 0 & J_{l} \\ -J_{l} & 0 \end{array}\label{defK}
 \right),\\
\nonumber D_l=\so(2l,\C), \,\so(l,l) & \qquad & K  =J_{2l}, 
\end{eqnarray}
where $ J_m=\left(\begin{array}{cccc}  &  &  & 1\\  &  & 1 &  \\ & \iddots & & \\ 1 &  &  &  \end{array}\right)\in \F^{m\times m}.$

The results for the series $A_l$, i.e. the classification of solvable Lie algebras with triangular nilradicals \cite{TW} can be briefly summed up as follows.

\begin{enumerate}
\item Every solvable Lie algebra $\s(n_{NR},q)$ with the triangular nilradical $\t(l+1)$ has the dimension 
$$d=q+n_{NR}, \qquad 1\leq q\leq l$$
where $n_{NR}=\frac{l(l+1)}{2}$ is the dimension of the nilradical $\t(l+1)$ and $l$ is the rank of the simple Lie algebra $A_l=\slmot(l+1,\F)$. 
\item A ``canonical basis'' $\{X^{\alpha},N_{ik}\},\ \alpha=1,\ldots,q,\ 1\leq i<k\leq l+1$ of $\s(n_{NR},q)$ exists in which the commutation relations are
\begin{eqnarray*}
{[N_{ik},N_{ab}]} & =& \delta_{ka}N_{ib}-\delta_{bi}N_{ak},\\
{[X^{\alpha},N_{ik}]}&=& \sum_{p<q} A^{\alpha}_{ik\,,\,pq}N_{pq},\\
{[X^{\alpha},X^{\beta}]}&=& \sigma^{\alpha\beta}N_{1(l+1)}.
\end{eqnarray*}
The matrices $A^{\alpha}$ are linearly nilindependent and upper triangular. For $q\geq 2$ they commute pairwise. The only off--diagonal matrix elements in $A^{\alpha}$ that may not vanish are
\begin{equation}\label{nonvanA}
A^{\alpha}_{{12\,,\,2(l+1)}},\quad A^{\alpha}_{{j(j+1)\,,\,1(l+1)}},\quad A^{\alpha}_{{l(l+1)\,,\,1l}},\quad 2\leq j \leq l-1
\end{equation}
The diagonal elements $A^{\alpha}_{{i(i+1)\,,\,i(i+1)}}$, $1\leq i \leq l$ are free  and determine the rest of the diagonal elements
$$ A^{\alpha}_{{ik\,,\,ik}}=\sum_{j=i}^{k-1} A^{\alpha}_{{j(j+1)\,,\,j(j+1)}}, \qquad i+1<k $$
\item All constants $\sigma^{\alpha\beta}$ vanish unless we have $A^{\gamma}_{1(l+1)\,,\,1(l+1)}=0\ $ for $\gamma=1,\ldots,q.$ The remaining off-diagonal elements $A^{\alpha}_{ik\,,\,ab}$ in equation (\ref{nonvanA}) also vanish, unless the diagonal elements satisfy $A^{\beta}_{ik\,,\,ik}=A^{\beta}_{ab\,,\,ab}\ $ for all $\beta=1,\ldots,q.$
\item The maximal value $q=l$ corresponds to exactly one solvable Lie algebra for which all matrices $A^{\alpha}$ are diagonal and all elements $X^{\alpha}$ commute. This algebra is isomorphic to the Borel subalgebra of $A_l$.
\item For the minimal value $q=1$ at most $l-1$ off-diagonal elements of $A^1$ are nonvanishing. They can be normalized to $1$ when $\F=\C$ and to $\pm 1 $ when $\F=\R$.
\end{enumerate}

In this article we shall show that essentially the same results hold for solvable Lie algebras with any Borel nilradical. The results for $A_l$ were proved using its explicit matrix realization. Here we shall prove analogous results for all simple Lie algebras simultaneously (the classical and exceptional complex ones and the split real ones) using bases for the nilradicals provided by the positive roots. This simultaneous treatment is made possible by the fact that all outer derivations of these nilradicals are known, due to the work of G.F. Leger and E.M. Luks \cite{LL}.\medskip

In Section \ref{prelim} we introduce some necessary terminology and notation.The general method of extending nilpotent Lie algebras to solvable ones using outer derivations is summed up in Section \ref{introsan}. Results on outer derivations of Borel nilradicals for simple Lie algebras are reviewed in Section \ref{odnbs}. Our main original results, formulated in three theorems, are presented in Section \ref{sebn}. Section \ref{concl} is devoted to conclusions.

\section{Notation}\label{prelim}
We shall need to refer to the Jacobi identity 
$$ [x,[y,z]]+ [y,[z,x]] + [z,[x,y]]=0$$
for a particular triplet $x,y,z$ of vectors in $\g$. For brevity, we speak of the Jacobi identity $(x,y,z)$. A Lie bracket of two vector subspaces is defined to be the span
$$ [\h,\tilde \h]=\span \{ [x,\tilde x] | x \in \h, \tilde x \in \tilde \h \}.$$

We recall the definitions of solvable and nilpotent Lie algebras and introduce a notation for the series of ideals involved:
The \emph{derived series} 
$ \g = \g^{(0)} \supset \g^{(1)} \supset \ldots \supset \g^{(k)} \supset \ldots $ 
is defined recursively
$$ \g^{(0)}=\g, \qquad  \g^{(k)} = [\g^{(k-1)},\g^{(k-1)}], \ k\geq 1.$$
If the derived series terminates, i.e. there exists $k \in \N$ such that $\g^{(k)} = 0$, then $\g$ is a \emph{solvable Lie algebra}.

The \emph{lower central series} $ \g = \g^{1} \supset \g^{2} \supset \ldots \supset \g^{k} \supset \ldots $ 
is again defined recursively
$$ \g^{1}=\g, \qquad \g^{k} = [\g^{k-1},\g], \qquad  k\geq 2. $$
If the lower central series terminates, i.e. there exists $k \in \N$ such that $\g^{k} = 0$, then $\g$ is 
called a \emph{nilpotent Lie algebra}. By definition, a nilpotent Lie algebra is also solvable.

A \emph{derivation} of $\g$ is a linear map $D: \ \g \rightarrow \g $
such that for any pair $x,y$ of vectors in $\g$ 
\begin{equation}\label{deriv} 
D([x,y])=[D(x),y]+[x,D(y)]
\end{equation}
holds. If a vector $z \in \g$ exists  such that $ D = {\rm ad} (z),$ i.e. $D(x)=[z,x], \ \forall x \in \g,  $ the derivation $D$ is called an \emph{inner derivation}, any other one is an \emph{outer derivation}. The space of inner derivations is denoted $\inn (\g)$, of all derivations $\der(\g)$.

We denote by $\dotplus$ the direct sum of vector spaces.

\section{Solvable Lie algebras with a given nilradical}\label{introsan}

In this section we briefly review the construction of all solvable Lie algebras with the given nilradical. The method is originally due to G.M. Mubarakzyanov who used it to construct all solvable Lie algebras up to dimension 5 in \cite{Mub1,Mub2} and 6--dimensional solvable Lie algebras with 5--dimensional nilradical in \cite{Mub3}.

Any solvable Lie algebra $\s$ contains a unique maximal nilpotent ideal $\n={\NR}(\s)$, the \emph{nilradical} of $\s$. We assume that $\n$ is given. That is, in some basis $( e_1, \ldots, e_n )$ of $\n$ we are given the Lie brackets
\begin{equation}\label{nilkom}
[e_j,e_k] = \sum_{l=1}^{n} {N_{jk}}^l e_l.
\end{equation}

Let us consider an extension of the nilpotent algebra $\n$ to a solvable Lie algebra $\s$, $\s \supsetneq \n$ having $\n$ as its nilradical. We call any such $\s$ a \emph{solvable extension} of the nilpotent Lie algebra $\n$. By definition, any such solvable extension $\s$ is non--nilpotent. 

We can assume without loss of generality that the structure of $\s$ is expressed in terms of linearly independent vectors $f_1,\ldots,f_q\in \s$ added to the basis $( e_1, \ldots, e_n )$ so that together they form a basis of $\s$. The derived algebra of a solvable Lie algebra is contained in the nilradical (see \cite{Jac}), i.e.
\begin{equation}\label{ssinn}
[\s,\s] \subseteq\n.
\end{equation}
It follows that the Lie brackets in $\s$ take the form
\begin{eqnarray}\label{Agam1}
[f_a,e_j] & = & \sum_{k=1}^{n} (\hat D_a)^k_{j} e_k, \; 1 \leq a \leq q, \  1 \leq j \leq n, \\ 
\label{Agam2} [f_a,f_b] & = & \sum_{j=1}^{n}  \gamma^j_{ab} e_j, \; 1 \leq a<b \leq q.
\end{eqnarray}
The operators $\hat D_a$ expressed as matrices in the basis $( e_1, \ldots, e_n )$ represent $f_a$ in the adjoint representation of $\s$ restricted to the nilradical $\n$, 
$$\hat D_a = {\rm ad} (f_a)|_\n. $$
The operators $\hat D_a$ must be outer derivations, otherwise
$\n \dotplus {\rm span} \{ f_a \}$ would be a nilpotent ideal in $\s$ contradicting the maximality and uniqueness of the nilradical $\n$. In fact, the maximality of $\n$ implies that no nontrivial linear combination of the operators $\hat D_a$ can be a nilpotent operator, i.e. the outer derivations $\hat D_1, \ldots, \hat D_q$ must be \emph{linearly nilindependent}. Let us remark that many nilpotent Lie algebras of dimension 7 and higher do not have any non--nilpotent derivations at all, i.e. don't possess any solvable extensions. Such nilpotent algebras are called \emph{characteristically nilpotent}. For dimensions large enough (i.e. $\geq 8$) there  exist subsets of characteristically nilpotent algebras which are open in Zariski topology on the variety of nilpotent Lie algebras \cite{Kh,GKh,AC}. Thus characteristically nilpotent  algebras are in a sense typical  nilpotent Lie algebras in large dimensions.

In view of equation (\ref{ssinn}), the commutators $[\hat D_a,\hat D_b]$ must be inner derivations of $\n$. The structure constants $\gamma^j_{ab}$ in the Lie brackets (\ref{Agam2}) are determined up to elements in the center $\z$ of $\n$ by a consequence of equation (\ref{Agam2})
\begin{equation}\label{DaDbge}
[\hat D_a,\hat D_b] = \sum_{j=1}^{n}  \gamma^j_{ab} {\rm ad} ( e_j)|_\n.
\end{equation}
Finally, the consistency of $\gamma^j_{ab}$ and $\hat D_a$ is subject to the constraint
\begin{equation}\label{fafbfc}
\sum_{j=1}^{n} \left(\gamma^j_{ab} \hat D_c(e_j) +\gamma^j_{bc} \hat D_a(e_j)+\gamma^j_{ca} \hat D_b(e_j) \right)=0
\end{equation}
coming from the Jacobi identity $(f_a,f_b,f_c)$. We remark that the left hand side of the condition (\ref{fafbfc}) always takes values in the center of $\n$ because the derivations $\hat D_a$ and ${\rm ad} ( e_j)|_\n$ themselves satisfy the Jacobi identity. 

Different sets of derivations $\hat D_a$ and their accompanying constants $\gamma^j_{ab}$ may correspond to isomorphic Lie algebras. The equivalence between sets of derivations $\hat D_a$ is generated by the following transformations:
\begin{enumerate}
\item We may add any inner derivation to any $\hat D_a$,
\begin{equation}\label{tffre}
 \hat D'_a=\hat D_a + \sum_{j=1}^{n} r_a^j \, {\rm ad} (e_j)|_\n, \qquad r_a^j \in \F.
\end{equation}
\item We may simultaneously conjugate all $\hat D_a$ by an automorphism $\Phi$ of $\n$, 
\begin{equation}\label{chbasnil}
 \hat D'_a = \Phi \circ \hat D_a \circ \Phi^{-1}, \qquad \Phi \in \mathrm{Aut}(\n) \subseteq GL(n,\F).
\end{equation}
\item We can change the basis in the space $ \sp \{ \hat D_1,\ldots, \hat D_q \}$.
\end{enumerate}
The corresponding changes in $\gamma^j_{ab}$ are then computed from equation~(\ref{DaDbge}).

\section{Outer derivations of nilradicals of Borel subalgebras}\label{odnbs}

Let $\g$ be a simple complex Lie algebra, $\g_0$ its Cartan subalgebra, $l=\rank\, \g=\dim \g_0$. Let us denote by $\Delta$ the set of all roots, by $\Delta^+$ the set of all positive roots and by $\Delta^S=\{ \alpha_1,\ldots, \alpha_l\}$ the set of simple roots. Let $\g_\lambda$ denote the root subspace of the root $\lambda$. We use the notation $\langle , \rangle$ for the scalar product on the $\sp$ of $\Delta^S$ (over $\R$) induced by the Killing form on $\g$. Let  $S_\beta$ denote the Weyl reflection with respect to the root $\beta$,
$$ S_\beta(\alpha) = \alpha - 2 \frac{\langle \alpha,\beta \rangle}{\langle \beta,\beta \rangle} \beta, \qquad \alpha \in \Delta . $$ 

As recalled above every (semi)simple complex Lie algebra $\g$ contains a unique (up to isomorphisms) maximal solvable subalgebra, its \emph{Borel subalgebra} $\mathfrak{b}(\g)$. It contains the Cartan subalgebra and all positive root vectors
$$ \mathfrak{b}(\g)=\g_0 \dotplus  \left( \dotplus \{ \g_\lambda | \lambda\in\Delta^+ \}\right).$$
The well--known properties of root systems imply that the Borel subalgebra is indeed a solvable subalgebra of $\g$ with the nilradical 
$$ \mathrm{NR}(\mathfrak{b}(\g))=  \dotplus \{ \g_\lambda | \, \lambda\in\Delta^+ \}.$$
For the sake of brevity we shall call the nilpotent Lie algebra $\mathrm{NR}(\mathfrak{b}(\g))$ the \emph{Borel nilradical} (although it is not the nilradical of the simple Lie algebra $\g$).

It is always possible to realize the simple Lie algebra $\g$ in such a manner that its Borel subalgebra $\mathfrak{b}(\g)$ is represented by upper triangular matrices. For $\slmot(l+1,\C)$, $\mathfrak{b}(\slmot(l+1,\C))$ is simply the set of all traceless $(l+1) \times (l+1)$ upper triangular matrices. The same holds true also for the \emph{split}, i.e. maximally noncompact, \emph{real form} $\slmot(l+1,\R)$ of $\slmot(l+1,\C)$. All other series of classical complex Lie algebras and their split real forms can be realized by matrices $X\in\C^{n\times n}$ or $X\in\R^{n\times n}$  satisfying equation~(\ref{xkkxt})
$$ X K + K X^t=0 $$
with $K$ for $B_l,C_l,D_l$ defined in equation~(\ref{defK}).

The corresponding Borel subalgebra $\mathfrak{b}(\g)$ is represented by upper triangular matrices satisfying the condition~(\ref{xkkxt}). The nilradical $ \mathrm{NR}(\mathfrak{b}(\g))$ is represented by strictly upper triangular matrices satisfying the condition~(\ref{xkkxt}). 

Let $\alpha_i, \; i=1,\ldots,l=\rank \, \g$ denote the simple roots, $\Delta^S=\{ \alpha_i \}_{i=1}^l$  and let 
$$\g_m= \dotplus \{\g_\lambda \, | \, \lambda=\sum_{i=1}^{l} m_i \alpha_i, \; \sum_{i=1}^{l} m_i \geq m \}.$$
The root vectors $ e_\alpha$, $\alpha\in\Delta^S$ generate the entire $\mathrm{NR}(\mathfrak{b}(\g))=\dotplus \{ \g_\lambda | \, \lambda\in\Delta^+ \}$ through commutators
$$ [\g_\lambda,\g_\mu]=\g_{\lambda+\mu} \qquad \mathrm{whenever} \quad \lambda,\mu,\lambda+\mu\in\Delta^+$$
and this implies that the ideals in the lower central series of the nilradical $\mathrm{NR}(\mathfrak{b}(\g)) $ of the Borel subalgebra are 
$$\left(\mathrm{NR}(\mathfrak{b}(\g))\right)^m=\g_m.$$
The center $\z$ of $\mathrm{NR}(\mathfrak{b}(\g))$ is one--dimensional and is spanned by $e_\lambda$ where $\lambda$ is the \emph{highest root} of $\g$, i.e. the unique root such that no root of the form $\lambda+\alpha_j$ exists. The center $\z$ coincides with the last nonvanishing ideal in the lower central series.\medskip

All derivations of the nilradical $\n=\mathrm{NR}(\mathfrak{b}(\g))$ were found in \cite{LL} and the result is as follows. 

\begin{prop}
Let $\g$ be a complex simple Lie algebra of rank $l$, $\g_0$ its Cartan subalgebra, $\Delta^S=\{ \alpha_1,\ldots,\alpha_l\}$ the set of simple roots and $\n=\mathrm{NR}(\mathfrak{b}(\g))$. Then the algebra of derivations of the nilradical $\n=\mathrm{NR}(\mathfrak{b}(\g))$ of the Borel subalgebra of a complex simple Lie algebra $\g$ satisfies
\begin{itemize}
 \item $\der (\n)=\out(\n) \dotplus \inn (\n)$, 
 \item $\dim \out(\n)=2l$,
 \item $\out(\n)= \sp \{ D_i, \tilde{D}_i \, | i=1,\ldots,l\}$ where the derivations $D_i, \tilde{D}_i$ are defined as follows. The derivations $D_i$ act diagonally in the basis of $\n$ consisting of positive root vectors $e_\alpha,$ $\alpha\in\Delta^+$
$$ D_i(e_{\alpha}) = m_i e_{\alpha}, \qquad \alpha=\sum_{j=1}^{l} m_j \alpha_j\in\Delta^+.$$ 
$\tilde{D}_i$ are nilpotent outer derivations which act on simple root vectors as
\begin{eqnarray}\label{dprimei}
\tilde{D}_i(e_\beta) & = e_\gamma, \qquad \mathrm{where}\quad \gamma={S_{\alpha_i}(\lambda)}, \quad  & \mathrm {if} \; \beta=\alpha_i,\\
\nn & =0,  \qquad & \mathrm{if}\; \beta=\alpha_j, \; j\neq i
\end{eqnarray}
where $\lambda$ is the highest root of $\g$. The action of $\tilde{D}_i$ on $e_\alpha,$ $\alpha\in\Delta^+ \backslash \Delta^S$ follows from the definition of a derivation~(\ref{deriv}).
\end{itemize}
\end{prop}
For future reference, we list Dynkin diagrams and the highest roots corresponding to all simple Lie algebras in Table~\ref{Ddhr}. Values of the highest roots were taken from \cite{FSS}.

For the sake of brevity, we shall write ${S_{i}(\lambda)}$ instead of ${S_{\alpha_i}(\lambda)}$ and introduce nonnegative integer constants $s_i$
$$ {S_{i}(\lambda)}=\lambda-s_i \alpha_i. $$
We notice that for $\g=A_l$ only two constants $s_i$, namely $s_1$ and $s_l$, are nonvanishing and equal to one; for all other simple algebras only one $s_i$ is nonvanishing and turns out to be equal to 1 or 2, see Table~\ref{nonvansi}. We also list the dimensions of the nilradicals $\mathrm{NR}(\mathfrak{b}(\g))$ in Table~\ref{nonvansi}.

\begin{table}
\begin{description}
\item[$A_l$]\hskip4mm 
\begin{picture}(62,15)
	\put(0,0){\circle*{2}}
	\put(10,0){\circle*{2}}
	\put(20,0){\circle*{2}}
	\put(30,0){\circle*{2}}
	\put(40,0){\circle*{2}}	
	\put(0,0){\line(1,0){10}}
	\put(10,0){\line(1,0){10}}
	\put(23,0){$\ldots$}
	\put(30,0){\line(1,0){10}}	
	\put(-3,-6){$\alpha_1$}
	\put(7,-6){$\alpha_2$}
	\put(17,-6){$\alpha_3$}
	\put(27,-6){$\alpha_{l-1}$}
	\put(37,-6){$\alpha_{l}$}
\end{picture}
$\lambda=\sum_{j=1}^{l} \alpha_i$
\item[$B_l$]\hskip4mm 
\begin{picture}(62,15)
	\put(0,0){\circle*{2}}
	\put(10,0){\circle*{2}}
	\put(20,0){\circle*{2}}
	\put(30,0){\circle*{2}}
	\put(40,0){\circle*{2}}	
	\put(0,0){\line(1,0){10}}
	\put(13,0){$\ldots$}
	\put(20,0){\line(1,0){10}}
	\multiput(30,-.5)(0,1){2}{\line(1,0){10}}	
	\put(-3,-6){$\alpha_1$}
	\put(7,-6){$\alpha_2$}
	\put(17,-6){$\alpha_{l-2}$}
	\put(27,-6){$\alpha_{l-1}$}
	\put(37,-6){$\alpha_{l}$}
        \put(34.5,-.8){$\rangle$}
\end{picture}
$\lambda=\alpha_1 + 2\sum_{j=2}^{l} \alpha_i$
\item[$C_l$]\hskip4mm 
\begin{picture}(62,15)
	\put(0,0){\circle*{2}}
	\put(10,0){\circle*{2}}
	\put(20,0){\circle*{2}}
	\put(30,0){\circle*{2}}
	\put(40,0){\circle*{2}}	
	\put(0,0){\line(1,0){10}}
	\put(13,0){$\ldots$}
	\put(20,0){\line(1,0){10}}
	\multiput(30,-.5)(0,1){2}{\line(1,0){10}}	
	\put(-3,-6){$\alpha_1$}
	\put(7,-6){$\alpha_2$}
	\put(17,-6){$\alpha_{l-2}$}
	\put(27,-6){$\alpha_{l-1}$}
	\put(37,-6){$\alpha_{l}$}
        \put(34.5,-.8){$\langle$}
\end{picture}
$\lambda= 2\sum_{j=1}^{l-1} \alpha_i+\alpha_{l}$
\item[$D_l$]\hskip4mm 
\begin{picture}(62,20)
	\put(0,0){\circle*{2}}
	\put(10,0){\circle*{2}}
	\put(20,0){\circle*{2}}
	\put(30,0){\circle*{2}}
        \put(30,6){\circle*{2}}
	\put(40,0){\circle*{2}}	
	\put(0,0){\line(1,0){10}}
	\put(13,0){$\ldots$}
	\put(20,0){\line(1,0){10}}
	\put(30,0){\line(1,0){10}}
	\put(30,0){\line(0,1){6}}	
	\put(-3,-6){$\alpha_1$}
	\put(7,-6){$\alpha_2$}
	\put(17,-6){$\alpha_{l-3}$}
	\put(27,-6){$\alpha_{l-2}$}
	\put(33,5){$\alpha_{l-1}$}
	\put(37,-6){$\alpha_{l}$}
\end{picture}
$\lambda= \alpha_{1}+2\sum_{j=2}^{l-2} \alpha_i+\alpha_{l-1}+\alpha_{l}$
\item[$E_6$]\hskip4mm 
\begin{picture}(62,20)
	\put(0,0){\circle*{2}}
	\put(10,0){\circle*{2}}
	\put(20,0){\circle*{2}}
	\put(20,6){\circle*{2}}
	\put(30,0){\circle*{2}}
        \put(40,0){\circle*{2}}	
	\put(0,0){\line(1,0){10}}
	\put(10,0){\line(1,0){10}}
	\put(20,0){\line(1,0){10}}
	\put(30,0){\line(1,0){10}}
	\put(20,0){\line(0,1){6}}	
	\put(-3,-6){$\alpha_1$}
	\put(7,-6){$\alpha_2$}
	\put(17,-6){$\alpha_{3}$}
	\put(27,-6){$\alpha_{4}$}
	\put(23,5){$\alpha_{6}$}
	\put(37,-6){$\alpha_{5}$}
\end{picture}
$\lambda= \alpha_{1}+2 \alpha_2+3\alpha_3+2 \alpha_4+\alpha_5+2 \alpha_6$
\item[$E_7$]\hskip4mm 
\begin{picture}(62,20)
	\put(0,0){\circle*{2}}
	\put(10,0){\circle*{2}}
	\put(20,0){\circle*{2}}
	\put(20,6){\circle*{2}}
	\put(30,0){\circle*{2}}
        \put(40,0){\circle*{2}}	
	\put(50,0){\circle*{2}}
	\put(0,0){\line(1,0){10}}
	\put(10,0){\line(1,0){10}}
	\put(20,0){\line(1,0){10}}
	\put(30,0){\line(1,0){10}}
	\put(40,0){\line(1,0){10}}
	\put(20,0){\line(0,1){6}}	
	\put(-3,-6){$\alpha_1$}
	\put(7,-6){$\alpha_2$}
	\put(17,-6){$\alpha_{3}$}
	\put(27,-6){$\alpha_{4}$}
	\put(23,5){$\alpha_{7}$}
	\put(37,-6){$\alpha_{5}$}
	\put(47,-6){$\alpha_{6}$}
\end{picture}
$\lambda= 2\alpha_{1}+3 \alpha_2+4\alpha_3+3 \alpha_4+2\alpha_5+ \alpha_6+2 \alpha_7$
\item[$E_8$]\hskip4mm 
\begin{picture}(62,20)
	\put(0,0){\circle*{2}}
	\put(10,0){\circle*{2}}
	\put(20,0){\circle*{2}}
	\put(20,6){\circle*{2}}
	\put(30,0){\circle*{2}}
        \put(40,0){\circle*{2}}	
	\put(50,0){\circle*{2}}
	\put(58,0){\circle*{2}}
	\put(0,0){\line(1,0){10}}
	\put(10,0){\line(1,0){10}}
	\put(20,0){\line(1,0){10}}
	\put(30,0){\line(1,0){10}}
	\put(40,0){\line(1,0){10}}
	\put(50,0){\line(1,0){8}}
	\put(20,0){\line(0,1){6}}	
	\put(-3,-6){$\alpha_1$}
	\put(7,-6){$\alpha_2$}
	\put(17,-6){$\alpha_{3}$}
	\put(27,-6){$\alpha_{4}$}
	\put(23,5){$\alpha_{8}$}
	\put(37,-6){$\alpha_{5}$}
	\put(47,-6){$\alpha_{6}$}
	\put(55,-6){$\alpha_{7}$}
\end{picture}
$\lambda= 2\alpha_{1}+4 \alpha_2+6\alpha_3+5 \alpha_4+4\alpha_5+ 3\alpha_6+2 \alpha_7+3 \alpha_8$
\item[$F_4$]\hskip4mm 
\begin{picture}(62,20)
\put(0,0){\circle*{2}} 	\put(10,0){\circle*{2}}	\put(20,0){\circle*{2}}	\put(30,0){\circle*{2}}	
\put(0,0){\line(1,0){10}} \multiput(10,-.5)(0,1){2}{\line(1,0){10}} \put(20,0){\line(1,0){10}}	
\put(-3,-4){$\alpha_1$}	\put(7,-4){$\alpha_2$}	\put(17,-4){$\alpha_3$}	\put(27,-4){$\alpha_4$}	
\put(14.5,-.8){$\rangle$}
\end{picture}
$\lambda= 2\alpha_{1}+3 \alpha_2+4\alpha_3+2 \alpha_4$
\item[$G_2$]\hskip4mm 
\begin{picture}(62,20)
\put(0,0){\circle*{2}} 	\put(10,0){\circle*{2}}		
\multiput(0,-1)(0,1){3}{\line(1,0){10}} 	
\put(-3,-4){$\alpha_1$}	\put(7,-4){$\alpha_2$}		
\put(4.5,-.8){$\rangle$}
\end{picture}
$\lambda= 2\alpha_{1}+3 \alpha_2$
\end{description}
\caption{Dynkin diagrams and highest roots of simple Lie algebras}\label{Ddhr}
\end{table}

\begin{table}
\begin{eqnarray*}
A_l \qquad & \qquad\dim NR(\b(A_l))= \frac{l(l+1)}{2} \qquad & \qquad s_1=s_l=1,\\
B_l \qquad & \qquad\dim NR(\b(B_l))= l^2 \qquad & \qquad s_2=1,\\
C_l \qquad & \qquad\dim NR(\b(C_l))= l^2 \qquad & \qquad s_1=2,\\
D_l \qquad & \qquad\dim NR(\b(D_l))= l(l-1) \qquad & \qquad s_2=1,\\
E_6 \qquad & \qquad\dim NR(\b(E_6))= 36\qquad & \qquad s_6=1,\\
E_7 \qquad & \qquad\dim NR(\b(E_7))= 63 \qquad & \qquad s_1=1,\\
E_8 \qquad & \qquad\dim NR(\b(E_8))=120 \qquad & \qquad s_7=1,\\
F_4 \qquad & \qquad\dim NR(\b(F_4))= 24 \qquad & \qquad s_1=1,\\
G_2 \qquad & \qquad\dim NR(\b(G_2))= 6\qquad & \qquad s_1=1.
\end{eqnarray*}
\caption{Dimensions of the nilradicals $NR(\b(\g))$ and nonvanishing constants $s_i$ in the equation ${S_{i}(\lambda)}=\lambda-s_i \alpha_i$}\label{nonvansi}\end{table}

Let us determine how $\tilde{D}_i$ act on the vectors $e_\beta$ when $\beta\in\Delta^+\backslash \Delta^S$. We proceed by induction, using the Leibniz property of derivations~(\ref{deriv}). Firstly, we evaluate
$$ \tilde{D}_i ([e_{\alpha_j},e_{\alpha_k}])= [\tilde{D}_i (e_{\alpha_j}),e_{\alpha_k}]+ [e_{\alpha_j},\tilde{D}_i (e_{\alpha_k})].$$
When both $i\neq j$ and $i\neq k$ the result is 0 by definition of $\tilde{D}_i$. Without loss of generality we take $i=j$ and $i\neq k$ and obtain
$$ \tilde{D}_i ([e_{\alpha_i},e_{\alpha_k}])= [\tilde{D}_i (e_{\alpha_i}),e_{\alpha_k}]=[e_{S_{i}(\lambda)},e_{\alpha_k}].$$
Were the result nonvanishing, we would have $S_{i}(\lambda)+\alpha_k\in\Delta^+$, i.e. also 
$$ S_{i}(S_{i}(\lambda)+\alpha_k)=\lambda+\alpha_k-2 \frac{\langle \alpha_k,\alpha_i \rangle}{\langle \alpha_i,\alpha_i \rangle} \alpha_i \in\Delta^+. $$ 
Since $2 \frac{\langle \alpha_k,\alpha_i \rangle}{\langle \alpha_i,\alpha_i \rangle}$ is an off--diagonal element of the Cartan matrix, it is a nonpositive integer. That means we would have a root larger than $\lambda$, contradicting the maximality of the highest root $\lambda$. Therefore, 
$$ \tilde{D}_i ([e_{\alpha_j},e_{\alpha_k}])=0$$
for any $j,k=1,\ldots,l$.
Next, let us assume that $\tilde{D}_i(e_{\beta})=0$ for all roots $\beta={\sum_{j=1}^l m_j \alpha_j}$ such that $2\leq \sum_{j=1}^l m_j \leq M$ and consider $ \tilde{D}_i ([e_{\alpha_j},e_{\beta}])$ with $\beta={\sum_{j=1}^l m_j \alpha_j},$ $\sum_{j=1}^l m_j= M$. When $j\neq i$ we again obtain $0$ using the Leibniz property~(\ref{deriv}). When $i=j$ we have
$$ \tilde{D}_i ([e_{\alpha_i},e_{\beta}])=[e_{S_{i}(\lambda)},e_{\beta}].$$
Using the Table~\ref{nonvansi} we find that for all simple algebras except $C_l$ we have either 
$$ S_{i}(\lambda)= \lambda-\alpha_i$$
or 
$$ S_{i}(\lambda)= \lambda.$$
Consequently, no root of the form $S_i(\lambda)+\beta$, where $\beta={\sum_{j=1}^l m_j \alpha_j}$, exists when $\sum_{j=1}^l m_j\geq 2$. Therefore the Lie bracket $[e_{S_{i}(\lambda)},e_{\beta}]$ vanishes.
In the special case of the algebras $C_l$ we have
$$ S_{1}(\lambda)= \lambda - 2 \alpha_1$$
and we must perform a more detailed analysis. Let us express $e_\beta$ as a multiple Lie bracket of simple root vectors $e_{\alpha_j}$,
$$ e_\beta=[e_{\alpha_{j_1}},\ldots,[e_{\alpha_{j_{M-1}}},e_{\alpha_{j_{M}}}]]$$
where $\beta=\sum_{K=1}^{M} \alpha_{j_K}$
and use the Jacobi identity sufficiently many times:
$$ [e_{S_{1}(\lambda)},e_\beta]= [[e_{S_{1}(\lambda)},e_{\alpha_{j_1}}],[\ldots,[e_{\alpha_{j_{M-1}}},e_{\alpha_{j_{M}}}]]]+\ldots+ [e_{\alpha_{j_1}},\ldots,[e_{\alpha_{j_{M-1}}},[e_{S_{1}(\lambda)},e_{\alpha_{j_{M}}}]]].$$ 
Whenever $j_{K}\neq1$ the corresponding commutator of $e_{\alpha_{j_{K}}}$ with $e_{S_{1}(\lambda)}$ vanishes. When $j_{K}=1$, we have
$$ [e_{S_{1}(\lambda)},e_{\alpha_1}] = c e_{\lambda-\alpha_1}$$
for some integer constant $c$. The only positive root vector which $e_{\lambda-\alpha_1}$ does not commute with is $e_{\alpha_1}$ so that any further commutator with any $e_{\alpha_j}$, $j\neq 1$ gives zero. Finally, the commutator of $e_{\lambda-\alpha_1}$ with $e_{\alpha_1}$ can in our computation arise only through
$$ [e_{S_{1}(\lambda)},[e_{\alpha_1},e_{\alpha_1}]]$$
which vanishes immediately. Therefore, in all cases we have shown that 
$$\tilde{D}_i([e_{\alpha_j},e_{\beta}])=0$$
 when $\beta={\sum_{j=1}^l m_j \alpha_j}$ and $\sum_{j=1}^l m_j = M$, concluding the induction step.

To sum up, we have just shown that for any simple complex Lie algebra $\g$ the derivations $\tilde{D}_i$ of the algebra $\mathrm{NR}(\mathfrak{b}(\g))$ give zero whenever they act on $e_\beta$, $\beta\in\Delta^+\backslash \Delta^S$. 

Let us assume from now on that $l>2$. Then we always have $S_{i}(\lambda)\notin \Delta^S$ for all $i=1,\ldots,l$ and consequently 
\begin{equation}\label{DiDje0}
\tilde{D}_i \circ \tilde{D}_j(e_{\alpha_k})=0
\end{equation}
for every $\alpha_k\in\Delta^S$. The previous calculation allows us to conclude that equation~(\ref{DiDje0}) must hold for any $\alpha\in\Delta^+$, i.e.
we have
$$ \tilde{D}_i \circ \tilde{D}_j=0$$
for any $i,j=1,\ldots,l$. The derivations $D_i$ obviously commute among each other and act diagonally on $\tilde{D}_j$, 
\begin{equation}\label{DitDjintDj}
 [D_i,\tilde{D}_j]\in \sp\{\tilde{D}_j\}.
\end{equation}

To conclude, we have just seen that under the assumption that $l$ is greater than $2$, the $2l$ outer derivations $D_i,\tilde{D}_i$ span a Lie subalgebra $\out(\mathrm{NR}(\mathfrak{b}(\g)))$ of the algebra of all derivations $\der(\mathrm{NR}(\mathfrak{b}(\g)))$. This algebra can be further decomposed into a semidirect sum of an $l$--dimensional Abelian ideal spanned by the nilpotent derivations $\tilde{D}_i$ and an $l$--dimensional Abelian subalgebra spanned by $D_i$. 

\section{Solvable extensions of the Borel nilradicals $\mathrm{NR}(\mathfrak{b}(\g))$}\label{sebn}
Let us now study the structure of any solvable Lie algebra with the nilradical $\mathrm{NR}(\mathfrak{b}(\g))$, $l=\mathrm{rank} \, \g>2$.

From the fact that there are only $l$ linearly nilindependent derivations $D_i$ in $\out(\mathrm{NR}(\mathfrak{b}(\g)))$ we conclude that the maximal number of nonnilpotent basis elements in any solvable Lie algebra $\s$ with the nilradical $\mathrm{NR}(\mathfrak{b}(\g))$ is $l$. One algebra with this number of nonnilpotent basis elements is already known, namely the Borel subalgebra $\mathfrak{b}(\g)$ of the simple Lie algebra $\g$. The following argument will show that it is the only one (up to isomorphisms).

\subsection{Solvable extensions of the Borel nilradicals of maximal dimension}
Let us assume that we have a solvable Lie algebra $\s$ with the nilradical $\n=\mathrm{NR}(\mathfrak{b}(\g))$ and $l=\mathrm{rank}\, \g$ nonnilpotent basis elements $f_i$. They define $l$ outer linearly nilindependent derivations $\hat D^i$ such that $\hat D^i=\ad(f_i)|_\n$. Using the transformation~(\ref{tffre}) we may choose the basis vectors $f_i$ so that 
$$\hat D^i=D_i+ \sum_{j=1}^{l} \omega_j^i \tilde{D}_j$$
where $D_i, \tilde{D}_j $ are the derivations defined above. Because $\hat D^i$ lie in the subalgebra $\out(\mathrm{NR}(\mathfrak{b}(\g)))$ of $\der(\mathrm{NR}(\mathfrak{b}(\g)))$ and at the same time $[\hat D^i,\hat D^j]\in\inn(\mathrm{NR}(\mathfrak{b}(\g)))$ must hold, we find that
\begin{equation}\label{dhcom}
[\hat D^k,\hat D^j]=0. 
\end{equation}
This requirement together with equation~(\ref{DitDjintDj}) in turn implies
\begin{equation}\label{oDDoDD}
 \omega_i^j [D_k,\tilde{D}_i]+\omega_i^k [\tilde{D}_i,D_j]=0. 
\end{equation}
for every $i,j,k=1,\ldots,l$ such that $k\neq j$ (no summation over $i$). For any given $i$ we can find ${{\tilde{i}}}$ such that $[\tilde{D}_i,D_{\tilde{i}}]\neq 0$. Consequently, the value of $\omega_i^{\tilde{i}}$ together with the root system specifying the Lie brackets $[D_k,\tilde{D}_i]$ completely determines all  $\omega_i^j$ for $j\neq {\tilde{i}}$.
Altogether, we still have one undetermined parameter $\omega_i^{\tilde{i}}$ for each $i=1,\ldots,l$. We proceed to eliminate these parameters through a suitable choice of automorphism in equation~(\ref{chbasnil}). 

We have $\left(\tilde{D}_i\right)^2=0$ and consequently $\tilde\phi_i(t_i)\equiv\exp\left(t_i \tilde{D}_i\right) = \mathbf{1}+t_i \tilde{D}_i$. Under the transformation~(\ref{chbasnil}) the derivations $D_j,\tilde{D}_j,\hat D^j$ change as follows
\begin{eqnarray}
\nonumber D_j  & \rightarrow\qquad & D_j+t_i [\tilde{D}_i,D_j], \\
\tilde{D}_j  & \rightarrow & \tilde{D}_j, \label{Borelautom} \\
\nonumber  \hat D^j = D_j+ \sum_{k=1}^{l} \omega_k^j \tilde{D}_k & \rightarrow & D_j+ t_i [\tilde{D}_i,D_j]+\sum_{k=1}^{l} \omega_k^j \tilde{D}_k. 
\end{eqnarray}
Now we consider step by step each $i=1,\ldots,l$. We take $\hat D^{\tilde{i}}$ which transforms nontrivially under the transformation~(\ref{Borelautom}) by definition of ${\tilde{i}}$, due to $[\tilde{D}_i,D_{\tilde{i}}]\neq 0$. We use it to set $\omega_i^{\tilde{i}}=0$ after the transformation. Equation~(\ref{oDDoDD}) then implies that after the  transformation all $\omega_i^j=0$. The automorphisms $\tilde\phi_i(t_i)$ commute among themselves, i.e. we can independently set to zero the constants $\omega_i^j$ belonging to different values of $i$ without changing the others. Therefore we have found that our derivations $\hat D^j$ can be brought to the form
$$ \hat D^j = D_j $$
through a conjugation by a suitable automorphism $\tilde\Phi=\tilde\phi_1(t_1)\circ\ldots\circ\tilde\phi_l(t_l)$ of $\mathrm{NR}(\mathfrak{b}(\g))$.

As we have just seen, the action of the nonnilpotent basis vectors $f_i$ on the nilradical $\mathrm{NR}(\mathfrak{b}(\g))$ is the same as in the Borel algebra. In order to show the uniqueness of the maximal solvable extension of $\mathrm{NR}(\mathfrak{b}(\g))$ it remains to be shown that we can always accomplish
$$ [f_i,f_j]=0. $$

The kernel of the adjoint representation $\ad:\n\rightarrow \glmot(\n)$ of the nilradical $\n$  is the center $\z$ of the algebra $\n$ which in our case is spanned by $e_\lambda$.
Consequently we have
$$ [f_i,f_j]=\gamma_{ij} e_\lambda, \qquad \gamma_{ij}=-\gamma_{ji} $$
which is the preimage of the relation $[\ad(f_i)|_\n,\ad(f_j)|_\n]=0$ (cf. equation~(\ref{DaDbge})).
Let us write the highest root in terms of simple roots as 
$$\lambda=\sum_{j=1}^{l} \lambda_j {\alpha_j}, \, \lambda_j\in\N.$$
The Jacobi identity~$(f_i,f_j,f_k)$ implies that
\begin{equation}\label{lglglg0}
 \lambda_i \gamma_{jk}+ \lambda_j \gamma_{ki}+\lambda_k \gamma_{ij}=0.
\end{equation} 
Let us perform a transformation
$$ f_i\rightarrow f_i+ \tau_i e_\lambda $$
which induces a change of the constants $\gamma_{ij}$
$$ \gamma_{ij} \rightarrow \gamma_{ij} + \lambda_i \tau_j- \lambda_j \tau_i.$$
In this way we may set to zero for example all $\gamma_{1j}$, $j=2,\ldots,l$ by selecting $\tau_1=1$ and choosing $\tau_j$ so that
$\gamma_{1j} + \lambda_1 \tau_j- \lambda_j=0$
(recall that all $\lambda_j\geq 1$). After such a transformation, equation~(\ref{lglglg0}) becomes
$$ \lambda_1 \gamma_{jk}=0, \qquad j,k\neq 1$$
and implies that all $\gamma_{jk}$ must now vanish. 
Therefore, we have found a basis $(e_\alpha,f_i|\alpha\in\Delta^+,i=1,\ldots,l)$ in our solvable algebra such that
\begin{eqnarray*}
[f_i,e_{\beta}] & = & m_i e_{\beta},\qquad \mathrm{where } \; \beta=\sum_{j=1}^l m_j \alpha_j,\\
 {[f_i,f_j]} & = & 0.
\end{eqnarray*}

To sum up, we have found that for any complex simple Lie algebra $\g$ such that $\mathrm{rank} \, \g>2$ the maximal solvable Lie algebra with the nilradical $\mathrm{NR}(\mathfrak{b}(\g))$ is unique and isomorphic to the Borel subalgebra $\mathfrak{b}(\g)$ of $\g$.

We notice that the same is true also when $\mathrm{rank} \, \g=1$ or $\mathrm{rank} \, \g=2$, i.e. $\g=\slmot(2),\slmot(3),\so(5)$ or $G_2$. In the case of $\g=G_2$ the derivation above can be taken over without modifications, since $S_{i}(\lambda)\in \Delta^+\backslash \Delta^S$ for all simple roots $\alpha_i$. In the cases of  $\g=\slmot(2),\slmot(3),\so(5)$ this is no longer true but the result follows from the investigation of Abelian nilradicals ($\slmot(2)$) \cite{NW}, Heisenberg nilradicals ($\slmot(3)$) \cite{RW}, and the nilradical $n_{4,1}$ of \cite{SW} ($\so(5)$).

Thus, we have proven the following theorem:
\begin{theorem}\label{csLaBsth}
Let $\g$ be a complex simple Lie algebra, $\mathfrak{b}(\g)$ its Borel subalgebra and $\n=\mathrm{NR}(\mathfrak{b}(\g))$ the nilradical of $\mathfrak{b}(\g)$. The solvable Lie algebra with the nilradical $\mathrm{NR}(\mathfrak{b}(\g))$ of the maximal dimension $\dim \n+\rank\, \g$ is unique and isomorphic to the Borel subalgebra $\mathfrak{b}(\g)$ of $\g$.
\end{theorem}

\subsection{Solvable extensions of the split real form of the nilradical $\mathrm{NR}(\mathfrak{b}(\g))$}\label{sesrf}
As stated in the Introduction, the maximal solvable subalgebra of a real form of a complex Lie algebra is not necessarily unique. Let us consider the split real form $\g$ and its maximal solvable subalgebra that consists of the maximally noncompact Cartan subalgebra (that remains diagonalizable over $\R$) and all positive root spaces. That means that it is spanned (over $\R$) by all $h_\alpha$ and $e_\alpha$ in the Weyl--Chevalley basis where $\alpha\in\Delta^+$. Let us call it the split Borel subalgebra and also denote it by $\mathfrak{b}(\g)$. The results of \cite{LL} also apply to real split Borel subalgebras.

Our Theorem~\ref{csLaBsth} holds for the split real form of a complex simple Lie algebra $\g$ with one exception: when $\g=\slmot(3,\R)$ there are several such maximal solvable algebras (see \cite{RW}). The proof for the cases $\rank \, \g>2$ or $\g=G_2$ is the same as above, the case $\g=\so(3,2)$ (the split real form of $\so(5,\C)$) follows from our paper \cite{SW}.

\subsection{Solvable extensions of the Borel nilradicals of smaller dimension}
After investigating the case of the maximal dimension solvable extension in previous Subsections let us consider the cases where the number of nonnilpotent elements of $\s$ is smaller than the rank of $\g$, i.e. $q<l$.

In such a case, we have derivations
\begin{equation}\label{hatdagen}
\hat D^a= \sum_{j=1}^{l} \left( \sigma^a_j D_j+ \omega_j^a \tilde{D}_j \right), \qquad a=1,\ldots,q  
\end{equation}
representing the elements $f_a$ in the adjoint representation of $\s$ on $\n$,  $ \hat D^a= \ad(f_a)|_\n$. The $q\times l$ matrix $\sigma=(\sigma^a_j)$ must have maximal rank, i.e. $q$, in view of the nilindependence of $\hat D^a$. However we can no longer set $\sigma^a_j$ equal to the Kronecker delta $\delta^a_j$ as was the case for $q=l$.

The condition (\ref{dhcom}), i.e. $[\hat D^a,\hat D^b]=0$, remains and implies
$$ \sum_{j,k=1}^{l} (\sigma^a_j \omega_k^b - \sigma^b_j \omega_k^a ) \left( \lambda_j - \delta_{jk} (1+s_j) \right) \tilde{D}_k=0 $$
which after separation of coefficients of linearly independent derivations $\tilde{D}_k$ gives a set of $\frac{q(q-1)}{2}l$ conditions
\begin{equation}\label{dhcomcoords}
 \sum_{j=1}^{l} (\sigma^a_j \omega_k^b - \sigma^b_j \omega_k^a ) \left( \lambda_j - \delta_{jk} (1+s_j) \right) =0, \qquad k=1,\ldots,l
\end{equation}
for $2q\cdot l$ unknown constants $\sigma^a_j,\omega_k^b$. When all $\omega_k^b=0$, any choice of $\sigma^a_j$ solves equation~(\ref{dhcomcoords}). 

Let us now find out how the constants $\sigma^a_j,\omega_j^a$  transform under transformations $\Phi$ of the form~(\ref{chbasnil}), i.e. a change of basis in the nilradical.  When $\Phi$ belongs to the connected component of $\mathrm{Aut}(\n)$ generated by exponentiation of $\der(\mathrm{NR}(\mathfrak{b}(\g)))$, the transformation~(\ref{chbasnil}) 
will not change the constants $\sigma^a_j$, owing to the fact that $[D_i,D_j]=0, \; [D_i,\tilde{D}_j]\in\sp \{ \tilde{D}_j \}$. (Certain permutations of the constants $\sigma^a_j$ may become possible under parity transformations belonging to other components of $\mathrm{Aut}(\n)$. E.g. the Dynkin diagram and the root system of $\slmot(l+1,\C)$ has the reflection symmetry $\alpha_j\leftrightarrow \alpha_{l+1-j}$ which gives rise to an automorphism of $\n$ interchanging $\sigma^a_j$ with $\sigma^a_{l+1-j}$.)  The constants $\omega_k^a$ transform as follows:
\begin{itemize}
 \item under automorphisms $\exp\left( u_j D_{j}\right)$ the constants $\omega_k^a$ get scaled by a factor $\exp\left( u_j d_{jk}\right)$ where $[D_j,\tilde{D}_k]=d_{jk} \tilde{D}_k$.
 \item under automorphisms $\tilde{\phi}_j(t_j)=\exp\left( t_j \tilde{D}_j\right)=\mathbf{1}+t_j \tilde{D}_j$ the constants $\omega_k^a$ are shifted to
\begin{equation}\label{omegaprime}
 \hat\omega_k^a= \omega_k^a-\left( \sum_{j=1}^{l} \sigma^a_j \lambda_j-\sigma^a_k (1+s_k) \right) t_k
\end{equation}
as follows from the evaluation of the commutator $[\tilde D_j,\sum_{k=1}^{l} \sigma^a_k D_k]$.
\end{itemize}
Let us assume that $\omega_k^a\neq 0$ for certain $a$ and $k$. We may add a suitable multiple of $\hat D^a$ to all $\hat D^b$, $b\neq a$ so that after the addition we have $\omega_k^b= 0$ for all $b\neq a$. Next, we check whether 
$ \sum_{j=1}^{l} \sigma^a_j \lambda_j-\sigma^a_k (1+s_k) \neq 0$. If it holds, we can use the transformation~(\ref{omegaprime}) to set also $\omega_k^a=0$. If not, we may attempt to modify $\hat D^a$ by addition of any multiple of $\hat D^b$, $b\neq a$.  If 
\begin{equation}\label{omegaprimecond}
 \sum_{j=1}^{l} \sigma^b_j \lambda_j-\sigma^b_k (1+s_k) \neq 0 
\end{equation}
holds for at least one $b=1,\ldots,q$, we can first change $\hat D^a$ and next eliminate $\omega_k^a=0$, i.e. now we have $\omega_k^b=0$ for all $b$, including $b=a$. Because the automorphisms $\tilde{\phi}_k(t_k)$ for different $k$ commute we may repeat the procedure for all values of $k$. Consequently, for any $k=1,\ldots,l$ we may set all $\omega_k^a=0$ whenever the condition~(\ref{omegaprimecond}) is satisfied for at least one $a\in\{1,\ldots,q\}$.

Let us find out for how many values of $k$ we can always transform the parameters $\omega_k^a$ into $\omega_k^a=0$. The matrix
$Q$ with entries $Q^{j}_{k}=\left( \lambda_j- (1+s_k)\delta_{jk} \right)$ has the form
$$ Q= \left( \begin{array}{ccccc}
           \lambda_1-(1+s_1) & \lambda_1 & \lambda_1 & \ldots & \lambda_1 \\
	   \lambda_2 & \lambda_2-(1+s_2) & \lambda_2 & \ldots & \lambda_2 \\
	   \lambda_3 & \lambda_3 & \lambda_3-(1+s_3) & \ldots & \lambda_3 \\
		\vdots & \vdots & & \ddots & \vdots \\
	   \lambda_l & \lambda_l & \lambda_l & \ldots & \lambda_l-(1+s_l) \\
          \end{array}
\right). $$
Subtracting the second column from the first, the third from the second etc. and using  $\lambda_k>0$ and $s_i\geq 0$ we find that it has always the maximal rank $l$, i.e.  $Q$ is a regular matrix. The expressions
 $R_k^a=\left( \sum_{j=1}^{l} \sigma^a_j \lambda_j-\sigma^a_k (1+s_k) \right)$ 
can be written as elements of the matrix product $R=\sigma\cdot Q$. Whenever the $k$-th column
$(R_k^a)_{a=1,\ldots,q}$ of the matrix $R$ is nonvanishing we can eliminate all parameters $\omega^a_k$ from our derivations $\hat D^a$. The number of indices $k$ such that the column $(R_k^a)_{a=1,\ldots,q}$ does not vanish is bounded from below by the rank of matrix $R$, i.e. also by the rank of matrix $\sigma$, which is equal to $q$. Therefore, we have at most $l-q$ values of the index $k$ such that some parameters $\omega^a_k$ are nonvanishing.

Now, after the simplifications introduced above, we revisit the commutativity condition~(\ref{dhcomcoords}). We observe that equation~(\ref{dhcomcoords}) can be written as a sum of terms, all of them containing the expression
\begin{equation}\label{slds}
\omega^d_k \sum_{j=1}^{l} \sigma^c_j \left( \lambda_j - \delta_{jk} (1+s_j) \right) 
\end{equation}
where $c=a$ and $d=b$ or vice versa. If the condition~(\ref{omegaprimecond}) is satisfied we have set $\omega^d_k=0$; otherwise, $\sum_{j=1}^{l} \sigma^c_j \left( \lambda_j - \delta_{jk} (1+s_j) \right)=0$. Either way, all terms of the form~(\ref{slds}) vanish and consequently the commutativity condition~(\ref{dhcomcoords}) is always satisfied.

Finally, we consider the Lie brackets between elements $f_a,f_b$. Equation~(\ref{Agam2}) takes the form
\begin{equation}\label{Agam2bn}
 [f_a,f_b]=\gamma_{ab} e_\lambda, \qquad a,b=1,\ldots,q.
\end{equation}
The constants $\gamma_{ab}$ in equation~(\ref{Agam2bn}) can be modified by transformations of the form
$$f_a\rightarrow f'_a=f_a+\tau_a e_\lambda.$$
Such a transformation does not change the derivations $\hat D_a$, $a=1,\ldots,q$ but may modify the Lie brackets $[f_a,f_b]$. Indeed, when there exists at least one $\hat D^a$ such that $\kappa=\sum_{j=1}^l \lambda^j \sigma^a_j\neq 0$, we have
$$ [f_a,e_\lambda]=\hat D^a(e_\lambda)=\kappa e_\lambda\neq 0.$$
Now we may choose $\tau_a=0$ and $\tau_b=-\gamma_{ab}/\kappa$, $a\neq b$  so that we obtain
$$ [f'_a,f'_b]=\gamma_{ab} e_\lambda - \frac{\gamma_{ab}}{\kappa} \kappa e_\lambda=0,$$
i.e. we have set $\gamma_{ab}=0$ for the given $a$ and all $b=1,\ldots,q$.
The Jacobi identity $(f_a,f_b,f_c)$ 
$$ \gamma_{ab} \hat D_c(e_\lambda)+\gamma_{bc} \hat D_a(e_\lambda)+\gamma_{ca} \hat D_b(e_\lambda)=0 $$
reduces to 
$$ \gamma_{bc} \hat D_a(e_\lambda)=\kappa \gamma_{bc} e_\lambda = 0, \qquad \forall\; b,c=2,\ldots,q $$
and implies that all constants $\gamma_{bc}=0$ vanish. Therefore, the basis elements $f_a$ in $\s$ can be chosen so that they commute whenever the condition
$$\sum_{j=1}^l \lambda^j \sigma^a_j\neq 0$$
holds for at least one $a$. On the other hand, if $\sum_{j=1}^l \lambda^j \sigma^a_j= 0$ for all $a$, the Jacobi identity $(f_a,f_b,f_c)$ is satisfied trivially.

Thus, we have proven the following theorem
\begin{theorem}\label{theononmaxcodim}
Let $\g$ be a complex simple Lie algebra or its split real form, $\b(\g)$ its (split) Borel subalgebra. Let $\g$ be different from $\slmot(3,\R)$. Any solvable extension $\s$ of the nilradical $NR(\b(\g))$ by $q$ nonnilpotent elements $f_a$, $a=1,\ldots,q\leq \rank \, \g$ is defined by $q$ commuting derivations $\hat D^a$ and a constant $q\times q$ antisymmetric matrix $\gamma=(\gamma_{ab})$.
The derivations $\hat D^a$ determine the Lie brackets
$$ [f_a,e_\alpha]=\hat D^a(e_\alpha), \qquad a=1,\ldots,q, \; \alpha\in \Delta^+ $$
and take the form
$$ \hat D^a= \ad(f_a)|_\n=\sum_{j=1}^{l} \left( \sigma^a_j D_j+ \omega_j^a \tilde{D}_j\right), \qquad a=1,\ldots,q,  $$
where the matrix $\sigma=(\sigma^a_j)$, $a=1,\ldots,q$, $j=1,\ldots,l$ has the maximal possible rank $q$.
For any given value of $k$ all parameters $\omega_k^a$ are equal to zero when the condition 
\begin{equation}\label{sls1s}
 \sum_{j=1}^{l} \sigma^a_j \lambda_j  -\sigma^a_k (1+s_k) \neq 0
\end{equation}
is satisfied for at least one $a\in\{1,\ldots,q\}$. The condition~(\ref{sls1s}) is always satisfied for at least $q$ values of the index $k$, i.e. there are at most $l-q$ values of $k$ such that some of the parameters $\omega_k^a$ are nonvanishing.

The matrix $\gamma=(\gamma_{ab})$ defines the Lie brackets
$$[f_a,f_b]=\gamma_{ab} e_\lambda, \qquad a,b= 1,\ldots,q.$$ 
When $$\sum_{j=1}^l \lambda^j \sigma^a_j\neq 0$$
holds for at least one $a\in\{1,\ldots,q\}$, the constants $\gamma_{ab}$ are all equal to $0$, i.e. 
$$ [f_a,f_b]=0.$$
\end{theorem}
Similarly as for Theorem~\ref{csLaBsth}, Theorem~\ref{theononmaxcodim} was proven above under the assumption that $\mathrm{rank} \, \g>2$. The derivation is the same when $\g=G_2$. When $\g=\slmot(2,\C)$ or $\g=\slmot(2,\R)$ the result is trivial. When $\g=\slmot(3,\C)$ the result follows from the consideration of Heisenberg nilradicals in \cite{RW}. When $\g=\slmot(3,\R)$ the results of Theorem~\ref{theononmaxcodim} do not hold as was already indicated in Subsection~\ref{sesrf} (for more details, see \cite{RW}). In the cases of  $\g=\so(5,\C)$ and $\g=\so(3,2)$ it follows from the investigation of the nilradical $n_{4,1}$ of \cite{SW}.

The conditions in Theorem~\ref{theononmaxcodim} are sufficient, i.e. any set of constants $\sigma^a_j,\omega_j^a$ and $\gamma_{ab}$ satisfying the properties listed in the theorem gives rise to a solvable extension of the nilradical $NR(\b(\g))$. On the other hand, the description presented in Theorem~\ref{theononmaxcodim} is not unique, i.e. different choices of $\sigma^a_j,\omega_j^a$ and $\gamma_{ab}$ may lead to isomorphic algebras. As already noted, we may replace the derivations $\hat D^a$ by any linearly independent combination of them thus changing all the parameters $\sigma^a_j,\omega_j^a$ and $\gamma_{ab}$. Also we may employ the scaling automorphisms to change the values of $\omega_j^a$ and $\gamma_{ab}$.

We remark that by virtue of indecomposability of the Borel nilradicals, all solvable Lie algebras described in Theorem \ref{theononmaxcodim} are indecomposable.

\subsection{Dimension $n_{NR}+1$ solvable extensions of the Borel nilradicals}
The results contained in Theorem~\ref{theononmaxcodim} can be further refined in the particular case $q=1$. Any single derivation $\hat D^1$ of the form~(\ref{hatdagen}) with nonvanishing vector $(\sigma^1_j)$ defines an $(n_{NR}+1)$--dimensional solvable extension of the nilradical $\n$. Let us investigate what restrictions we may impose on the parameters $\omega_k^1$ through a choice of a suitable transformation~(\ref{chbasnil}).

In the transformation~(\ref{omegaprime}) we choose $t_k$ so that $\hat\omega_k^1=0$ whenever the expression $ \sum_{j=1}^{l} \sigma^1_j \lambda_j-\sigma^1_k (1+s_k) $ is nonvanishing. Otherwise, the parameter $\omega_k^1$ remains nonvanishing after conjugation by any automorphism from $\exp(\der(\mathrm{NR}(\mathfrak{b}(\g))))$.

We already know that there can be at most $l-1$ nonvanishing parameters $\omega_k^1=0$. On the other hand, for any choice of up to $l-1$ indices $k_u, \; u=1,\ldots,l-1$ we can find constants $\sigma^1_j$ such that 
$$ \sum_{j=1}^{l} \sigma^1_j \lambda_j-\sigma^1_{k_u} (1+s_{k_u}) =0, \qquad u=1,\ldots,l-1$$
and consequently the parameters $\hat\omega_{k_u}^1$ cannot be set to zero by conjugation by any automorphism from the connected part of the automorphism group $\mathrm{Aut}(\n)$. 
That means that there are up to $l-1$ nondiagonal nonremovable parameters in the derivation $\hat D^1$ (whose positions may be changed by transformations belonging to other components of $\mathrm{Aut}(\n)$ if the root system of $\g$ allows them).

The nonvanishing parameters $\omega_k^1$ can be scaled to $1$ over the field of complex numbers. Over the field of real numbers, the usual problems with square roots may arise, so in some cases we can only normalize $\omega_k^1$ to $\omega_k^1=\pm 1$. 

In order to show this, let us consider the automorphisms $\hat\phi_j(v_j)$, $v_j\neq 0$ defined by
$$\hat\phi_j(v_j) e_\alpha= v_j^{m_j} e_\alpha, \quad \alpha=\sum_{j=1}^{l} m_j \alpha_j\in\Delta^+ $$
which generalize the inner automorphisms $\phi_j(u_j)=\exp\left( u_j D_j\right)$ to include also scaling by negative numbers. Under the transformation $\hat\phi_j(v_j)$ the constant $\omega_k^1$ gets scaled by the factor $v_j^{d_{jk}}$ where $d_{jk}=\lambda_j - \delta_{jk}(1+s_k)$,
$$ \omega_k^1 \rightarrow v_j^{\lambda_j - \delta_{jk}(1+s_k)}   \omega_k^1.$$
Combining $l$ such commuting automorphisms together, 
\begin{equation}\label{hpv}
\hat\phi(v)=\hat\phi_1(v_1)\circ \ldots \circ \hat\phi_l(v_l),  
\end{equation}
we transform 
\begin{equation}\label{vluo}
 \omega_k^1 \rightarrow \frac{\Pi_{j=1}^{l} v_j^{\lambda_j}}{ v_k^{1+s_k}} \; \omega_k^1.
\end{equation}
Let us consider what can be accomplished using equation~(\ref{vluo}). Let us first deal with all $k$ such that $s_k=0$ (and $\omega_k^1\neq 0$ by assumption). We set $v_k=\omega_k^1$. We still have at our disposal at least one constant $v_i$ such that $s_i\neq 0$. We can use it to set $\Pi_{j=1}^{l} v_j^{\lambda_j}=1$
over the field of complex numbers. Over the field of real numbers we can set $\Pi_{j=1}^{l} v_j^{\lambda_j}=1$ when  $\lambda_i$ is odd and $\Pi_{j=1}^{l} v_j^{\lambda_j}=\pm 1$ when $\lambda_i$ is even. Thus all nonvanishing constants $\omega_k^1$ such that $s_k=0$ were brought by a suitable automorphism~(\ref{hpv}) to $1$ over the field of complex numbers and to $\epsilon=\pm 1$ over the field of real numbers (where the value of $\epsilon$ is the same for all $k$). 

Assuming that the above described rescaling was already performed, we want to scale $\omega_i^1$ with $s_i\neq 0$ to a convenient value. We recall that at most $l-1$ parameters $\omega_j^1$ are nonvanishing. Therefore, if $\omega_i^1\neq 0$ we have at least one $k$, $k\neq i$ such that $\omega_k^1=0$. We choose an automorphism  $\hat\phi(v)$ such that $v_j=1$ whenever $j\neq i,k$, 
$$v_i^{1+s_i}={\omega_i^1}$$
and $v_k$ is chosen so that $\Pi_{j=1}^{l} v_j^{\lambda_j}=1$.  Over the field of real numbers we may accomplish only $v_i^{1+s_i}=\pm \frac{1}{\omega_i^1}$ when $s_i$ is odd and $\Pi_{j=1}^{l} v_j^{\lambda_j}=\pm 1$  when $\lambda_k$ is even. The automorphism $\hat\phi(v)$ constructed in this way does not affect the values $\omega_j^1$ set to $1$ in the previous step  (or scales them by a common minus factor when $\Pi_{j=1}^{l} v_j^{\lambda_j}=- 1$) and allows us to scale $\omega_i^1$ with $s_i\neq 0$ to $1$ ($\pm 1$, respectively). In the case of the algebra $A_l$ the procedure is repeated once again for the second index ${\hat{i}}$ such that $s_{\hat{i}}\neq 0$.

To sum up, we have the following theorem.

\begin{theorem} Let $\g$ and $\b(\g)$ satisfy the hypotheses of Theorem \ref{theononmaxcodim}.
Any solvable extension of the nilradical $NR(\b(\g))$ by one nonnilpotent element is up to isomorphism defined by a single derivation
$$ \hat D=\ad (f_1)|_\n = \sum_{j=1}^{l} \left( \sigma_j D_j+ \omega_j \tilde{D}_j \right) $$
chosen so that the first nonvanishing parameter $\sigma_j$ is equal to one.  The parameter $\omega_k$ vanishes whenever $ \sum_{j=1}^{l} \sigma_j \lambda_j-\sigma_k (1+s_k) \neq 0$. At most $l-1$ parameters $\omega_k$ are nonvanishing. They are all equal to $1$ over the field of complex numbers. Over the field of real numbers they are equal to $\pm 1$ and all parameters $\omega_k$ with $s_k=0$ have the same sign.
\end{theorem}

\section{Conclusions}\label{concl}
The need for a classification of solvable Lie algebras of higher dimensions in physics was discussed in our previous articles \cite{SW,SW1}. This arises in particular in the classification of higher dimensional Einstein spaces, or other Riemannian or pseudo--Riemannian spaces that can occur in string theories, brane cosmology and other elementary particle theories \cite{Petrov,KSHMC,Landau,Bianchi,Tur3,GSW,BBS,RS}.

Even if no complete classification of solvable Lie algebras exists, once a Lie algebra $\g$ occurs in a physical problem its radical $R(\g)$ can be identified. In turn the nilradical $NR(\g)$ can be obtained \cite{RWZ}.
If $NR(\g)$ belongs to one of the series that have been extended to solvable Lie algebras, then the complete identification of $\g$ is achieved.

Algorithms for calculating $R(\g)$ and $NR(\g)$ exist \cite{RWZ}. Their effectiveness of course depends on the dimension of the considered Lie algebra. They have been recently computerized and are available as a package for computer algebra system Maple \cite{AT}. (For older programs in PASCAL see \cite{RWZalg1,RWZalg2,RWZalg3}.)

\ack The research of L\v{S} was supported by the research plan MSM6840770039 of the Ministry of Education of the Czech Republic. The research of PW was partly supported by a research grant from NSERC of Canada. L\v{S} thanks the Centre de recherches math\'ematiques, Universit\'e de Montr\'eal for hospitality during work on the manuscript. We thank professors I. Anderson, A.G. Elashvili and J. Patera for interesting and helpful discussions.

\end{document}